\documentclass[a4paper,twoside,12pt]{article}
\input{obsjkt.sty}

\newcommand{\Msun}{\ensuremath{\,{\rm M}_\odot}}                  % Solar mass symbol
\newcommand{\Rsun}{\ensuremath{\,{\rm R}_\odot}}                  % Solar radius symbol
\newcommand{\rhosun}{\ensuremath{\,\rho_\odot}}                   % Solar density symbol
\newcommand{\Teff}{\ensuremath{T_{\rm eff}}}                      % Effective temperature symbol
\newcommand{\degr}{\ensuremath{^\circ}}                           % Degree symbol
\renewcommand{\kms}{\,km\,s$^{-1}$}                               % km/s symbol
\newcommand{\etal}{\textit{et al.}}                               % et al. in italics

\newcommand{\Msunnom}{\hbox{$\mathcal{M}^{\rm N}_\odot$}}
\newcommand{\Rsunnom}{\hbox{$\mathcal{R}^{\rm N}_\odot$}}
\newcommand{\Lsunnom}{\hbox{$\mathcal{L}^{\rm N}_\odot$}}

\OBSheader{Rediscussion of eclipsing binaries: KX Cnc}{J.\ Southworth}{2021 Feb}

\begin{document} %%%%%%%%%%%%%%%%%%%%%%%%%%%%%%%%%%%%%%%%%%%%%%%%%%%%%%%%%%%%%%%%%%%%%%%%%%%%%%%%%%%%%%%%%%%%%%%%%%%%%%%%%%%%%%%%%%%%%%%%%%%%%%%%%%%%
%%%%%%%%%%%%%%%%%%%%%%%%%%%%%%%%%%%%%%%%%%%%%%%%%%%%%%%%%%%%%%%%%%%%%%%%%%%%%%%%%%%%%%%%%%%%%%%%%%%%%%%%%%%%%%%%%%%%%%%%%%%%%%%%%%%%%%%%%%%%%%%%%%%%%

\OBStitle{Rediscussion of eclipsing binaries. Paper II. \\ The eccentric solar-type system KX\,Cancri}

\OBSauth{John Southworth}

\OBSinstone{Astrophysics Group, Keele University, Staffordshire, ST5 5BG, UK}

\OBSabstract{KX\,Cancri is an eclipsing binary containing two G-type stars with an orbital period of 31.2\,d and an eccentricity of 0.47. These qualities make it a promising candidate for a benchmark solar-type binary system. We analyse the first light curve of this system to have complete coverage of both primary and secondary eclipses, obtained using the Transiting Exoplanet Survey Satellite (TESS). We augment these data with published radial velocities and measure the masses to be $1.134 \pm 0.003$\Msun\ and $1.124 \pm 0.005$\Msun\ and the radii to be $1.053 \pm 0.006$\Rsun\ and $1.059 \pm 0.005$\Rsun. A ratio of the radii near unity is strongly preferred by the TESS data, in contrast to existing ground-based light curves. The distance to the system measured from the radii and \Teff\ values of the stars agrees well with the trigonometric parallax from the \textit{Gaia} satellite. The properties of the system are consistent with theoretical predictions for a super-solar metallicity and an age of 1.0--1.5\,Gyr. A detailed analysis of the photospheric properties of the stars based on high-resolution spectra is encouraged.}

%%%%%%%%%%%%%%%%%%%%%%%%%%%%%%%%%%%%%%%%%%%%%%%%%%%%%%%%%%%%%%%%%%%%%%%%%%%%%%%%%%%%%%%%%%%%%%%%%%%%%%%%%%%%%%%%%%%%%%%%%%%%%%%%%%%%%%%%%%%%%%%%%%%%%

\section*{Introduction}

Eclipsing binary stars provide our main source of direct measurements of the physical properties (mass, radius, luminosity) of normal stars \cite{Andersen91aarv,Torres++10aarv}. Their properties can be measured using only observational data and algebra \cite{Russell12apj,Hilditch01book} so are valuable in calibrating and assessing our understanding of stellar physics \cite{Pols+97mn,Chen+14mn,ClaretTorres18apj}, the chemistry of the universe \cite{PaczynskiSienkiewicz84apj,Ribas+00mn}, and the cosmological distance scale \cite{Pietrzynski+19nat,Freedman+20apj}.

Arguably the most important class of eclipsing system is that of the detached eclipsing binaries (dEBs), because their component stars have experienced no mass transfer so are representative of normal stars. The best benchmark system has well-separated stars so tidal effects are negligible, and precise measurements of the masses, radii, effective temperature (\Teff) values, luminosities, and photospheric chemical abundances of the two components.

dEBs containing stars similar to our Sun are particularly useful because they allow a direct comparison with the star by far the best-understood by humans, and thus aid the understanding of stellar structure as a function of time and chemical composition around the solar fiducial point. Those with an orbital period longer than approximately 10\,d are negligibly affected by tides and thus are most directly comparable to single stars such as our Sun; examples of such systems are V1094\,Tauri \cite{Maxted+15aa}, LL\,Aquarii \cite{Me13aa,Graczyk+16aa}, Kepler-34 \cite{Welsh+12nat} and KIC\,7177553\,S \cite{Lehmann+16apj}.

This is the second of a series of papers aimed at providing improved measurements of the physical properties of dEBs using data that have recently become available in photometric surveys performed by space telescopes \cite{Kirk+16aj,Deleuil+18aa,Ricker+15jatis}. A particular aim is curation of the DEBCat\footnote{\texttt{https://www.astro.keele.ac.uk/jkt/debcat/}} (Detached Eclipsing Binary Catalogue) list of dEBs with precise mass and radius measurements \cite{Me15debcat}. A detailed justification is given in the first paper of the series \cite{Me20obs}, which presented a reanalysis of the bright B-type dEB $\zeta$\,Phoenicis.

\begin{table}[t]
\caption{\em Basic information on KX\,Cnc \label{tab:info}}
\centering
\begin{tabular}{lll}
{\em Property}                      & {\em Value}            & {\em Reference}                   \\[3pt]
Henry Draper designation            & HD 74057               & \cite{CannonPickering19anhar}     \\
\textit{Hipparcos} designation      & HIP 42753              & \cite{Hip97}                      \\
\textit{Gaia} DR2 ID                & 709910784966516992     & \cite{Gaia18aa}                   \\
\textit{Gaia} parallax              & $20.282 \pm 0.051$ mas & \cite{Gaia18aa}                   \\
$B$ magnitude                       & $7.76 \pm 0.01$        & \cite{Hog+00aa}                   \\
$V$ magnitude                       & $7.19 \pm 0.01$        & \cite{Hog+00aa}                   \\
$J$ magnitude                       & $6.509 \pm 0.021$      & \cite{Cutri+03book}               \\
$H$ magnitude                       & $6.278 \pm 0.027$      & \cite{Cutri+03book}               \\
$K_s$ magnitude                     & $6.213 \pm 0.018$      & \cite{Cutri+03book}               \\
Spectral type                       & G0\,V + G1\,V          & This work                         \\[10pt]
\end{tabular}
\end{table}

In this work we analyse the dEB KX\,Cancri (HD\,74057), which contains two solar-type stars on a relatively long-period orbit so is an excellent candidate for becoming a benchmark system. Basic observational properties of the system are given in Table\,\ref{tab:info}. KX\,Cnc was discovered to be eclipsing by Davies \cite{Davies06pz,Davies07pz} and independently by Sowell \etal\ \cite{Sowell++12aj}. The latter work presented extensive photometry in the Str\"omgren $b$ and $y$ passbands plus radial velocities (RVs) for the two stars from a total of 26 high-resolution coud\'e spectra. They determined the masses and radii to precisions of 0.3\% and 0.2\%, respectively. Such a precision in radius is surprising, as the system does not show total eclipses and the first and last contact points of the secondary eclipse were not observed. However, a space-based light curve of this system, with full coverage of both primary and secondary eclipses, is now available. The analysis of KX\,Cnc using these new data is described below.

%%%%%%%%%%%%%%%%%%%%%%%%%%%%%%%%%%%%%%%%%%%%%%%%%%%%%%%%%%%%%%%%%%%%%%%%%%%%%%%%%%%%%%%%%%%%%%%%%%%%%%%%%%%%%%%%%%%%%%%%%%%%%%%%%%%%%%%%%%%%%%%%%%%%%

\section*{Observational material}

\begin{figure}[t] \centering \includegraphics[width=\textwidth]{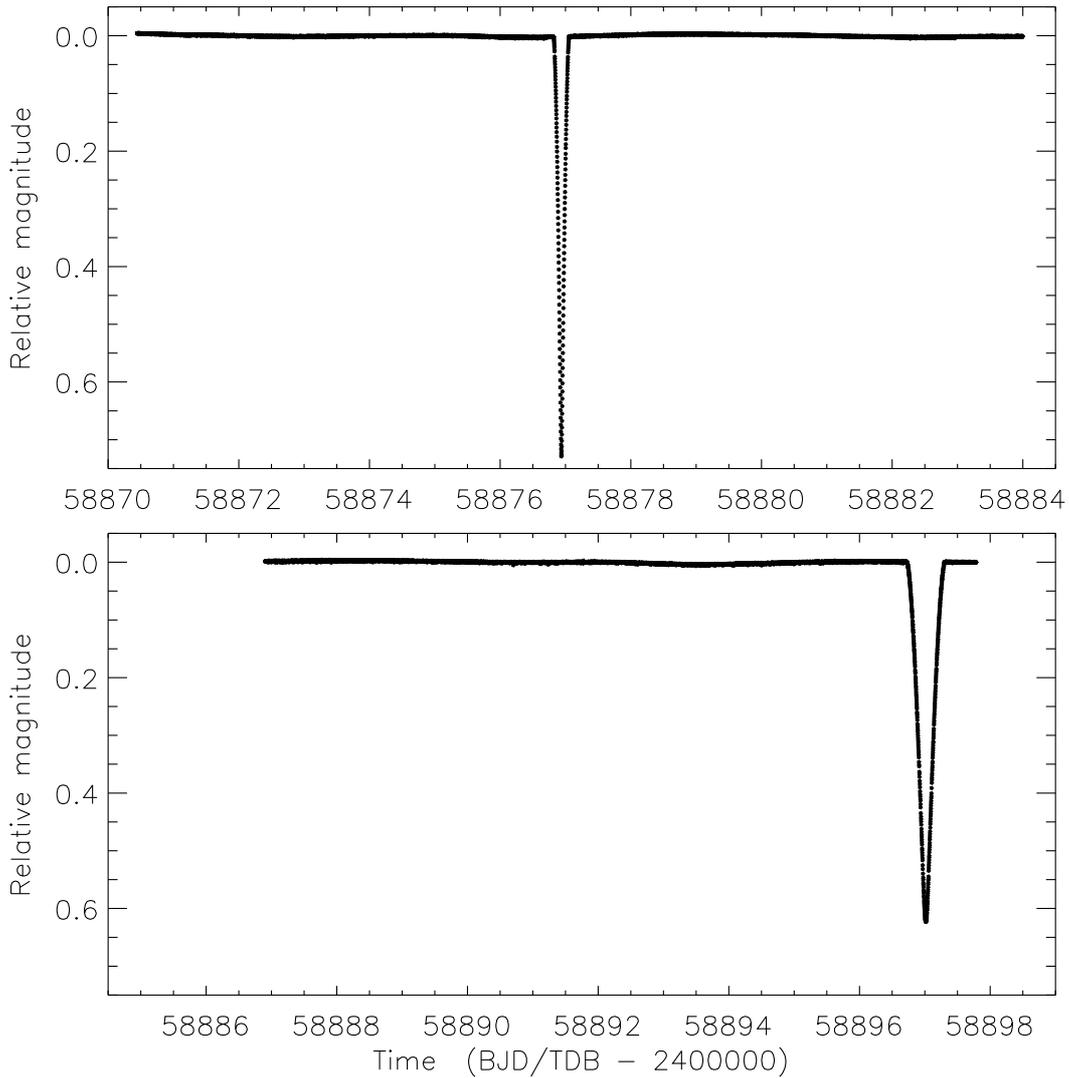} \\
\caption{\label{fig:time} TESS simple aperture photometry of KX\,Cnc. The upper and
lower plots show the observations either side of the mid-sector pause for data download.}
\end{figure}

As with Paper\,I \cite{Me20obs} the new data for the target dEB come from the NASA TESS satellite \cite{Ricker+15jatis}. KX\,Cnc was observed in camera 1 during Sector 21 (2020/01/21 to 2020/02/18). The light curve covers 27.4\,d, with a break near the midpoint for download of data to Earth, at a cadence of 120\,s.

For this work we used the simple aperture photometry (SAP) and not the pre-search data conditioning (PDC) light curve \cite{Jenkins+16spie}. Our experience of TESS data is that the PDC light curves, which receive additional processing beyond that for the SAP data, can become unreliable when there is strong variability in the target star (e.g.\ deep eclipses such as found in KX\,Cnc).

We retained only data with no flagged problems (QUALITY $=$ 0), comprising 17\,327 datapoints. The data were further trimmed by removing points more than 1.5 eclipse durations from the midpoint of an eclipse, as the out-of-eclipse data are essentially devoid of information on the masses and radii of the stars, leaving a total of 1618 datapoints. We ignored the errorbars of the measurements, as they are far too small.

%%%%%%%%%%%%%%%%%%%%%%%%%%%%%%%%%%%%%%%%%%%%%%%%%%%%%%%%%%%%%%%%%%%%%%%%%%%%%%%%%%%%%%%%%%%%%%%%%%%%%%%%%%%%%%%%%%%%%%%%%%%%%%%%%%%%%%%%%%%%%%%%%%%%%

\section*{Analysis of ground-based light curves}

\begin{figure}[t] \centering \includegraphics[width=\textwidth]{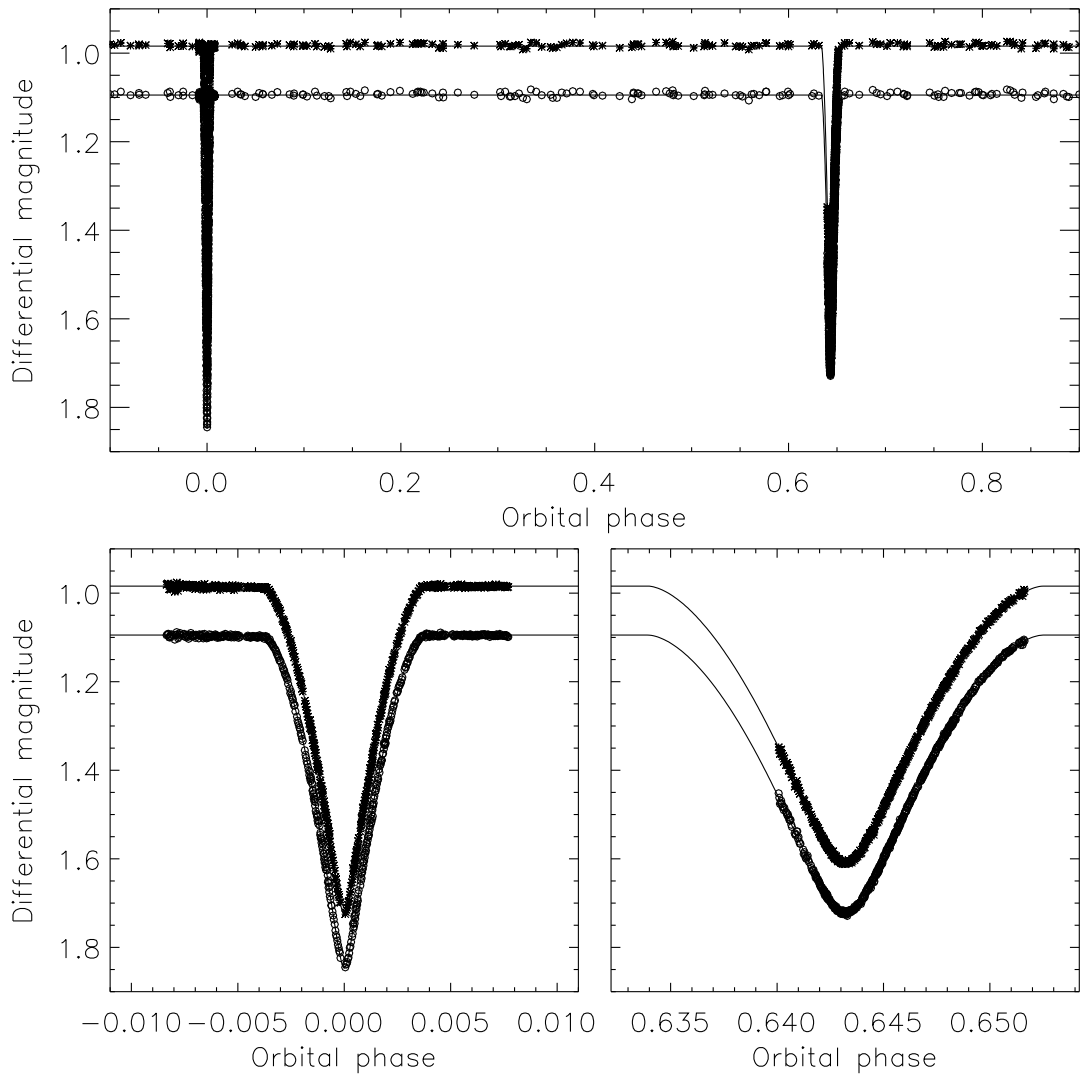} \\
\caption{\label{fig:sowell} The Str\"omgren $b$-band (crosses) and $y$-band (open
circles) data obtained Sowell \etal\ \cite{Sowell++12aj}, versus the \textsc{jktebop}
best fits from the current work (solid lines). The full light curve is shown in the
upper panel, and close-ups of the primary and secondary eclipses in the lower panels.}
\end{figure}

\begin{figure}[t] \centering \includegraphics[width=\textwidth]{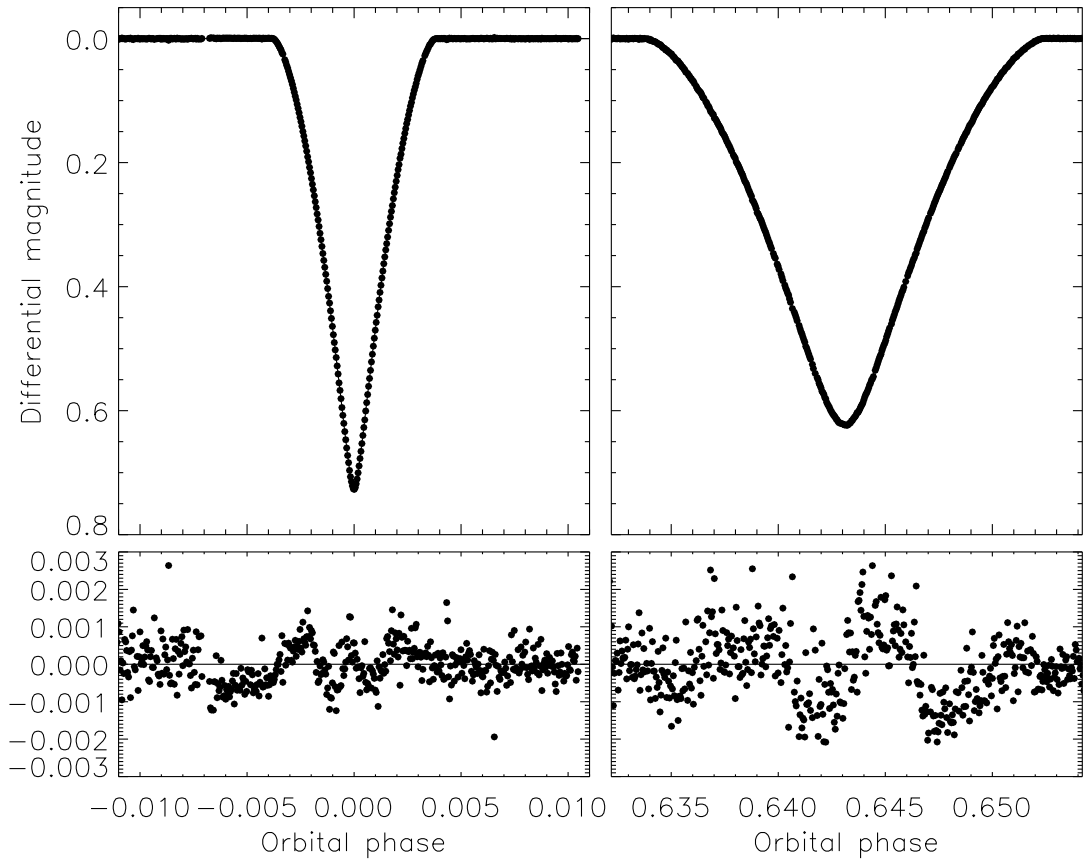} \\
\caption{\label{fig:tess} The TESS light curve of KX\,Cnc around the primary (left)
and secondary (right) eclipses. The {\sc jktebop} best fit is shown using a solid line.
The lower panels show the residuals of the fit on a magnified scale.} \end{figure}

\begin{figure}[t] \centering \includegraphics[width=\textwidth]{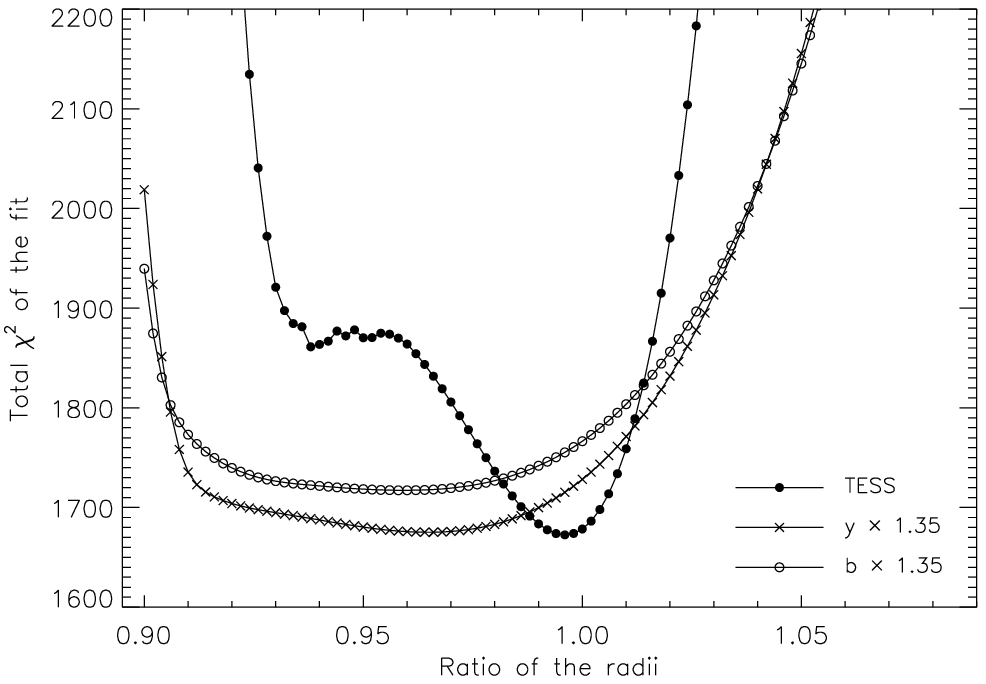} \\
\caption{\label{fig:kfix} The total $\chi^2$ of the fit to the three available light
curves of KX\,Cnc, for a grid of values of the ratio of the radii. The results for
each light curve are shown with different symbols (see key). Each fit also included
the RVs of the two stars from Sowell \etal\ \cite{Sowell++12aj}. The results for the
$y$ and $b$ light curves have been multiplied by a factor of 1.35 to give them
approximately the same minimum $\chi^2$ as the TESS data.} \end{figure}

The stars are well-separated and almost spherical, so the system can be reliably modelled using the {\sc jktebop} code\footnote{\texttt{http://www.astro.keele.ac.uk/jkt/codes/jktebop.html}}, for which we used version 40 \cite{Me08mn,Me13aa}. Results from the use of {\sc jktebop} have been found to be in good agreement with other codes for well-detached systems \cite{Maxted+20mn}. We follow the definition that the primary star is the one eclipsed during the deeper eclipse; we designate this as star A and the secondary star as star B. In the case of KX\,Cnc, star A has a larger mass and higher \Teff\ than star B, but not by signifciant amounts.

The radii of the stars in the {\sc jktebop} fits were parameterised by sum and ratio of the fractional radii ($r_{\rm A} = \frac{R_{\rm A}}{a}$ and $r_{\rm B} = \frac{R_{\rm B}}{a}$ where $R_{\rm A}$ and $R_{\rm B}$ are the true radii and $a$ is the orbital semimajor axis). The orbital shape was parameterised by the Poincar\'e elements ($e\cos\omega$ and $e\sin\omega$ where $e$ is the orbital eccentricity and $\omega$ is the argument of periastron). We included $r_{\rm A}+r_{\rm B}$, $k$, $e\cos\omega$, $e\sin\omega$, the orbital inclination and the central surface brightness ratio as fitted parameters. Limb darkening was represented using the quadratic law, the coefficients were assumed to be the same for both stars, the linear coefficient was fitted, and the nonlinear coefficient was fixed to a suitable value from Claret \cite{Claret00aa,Claret18aa}. These assumptions were checked and found to have a negligible impact on the best fits.

We began by modelling the photometry and RVs of the two stars from Sowell \etal\ \cite{Sowell++12aj}. We fitted the $b$-band and $y$-band data separately as {\sc jktebop} can only deal with one passband at once. Uncertainties were computed using Monte Carlo simulations \cite{Me08mn} after adjusting the data errors so that the reduced $\chi^2$ of each of the three datasets ($y$-band light curve and RVs of each star) was unity. The orbital period and reference time of minimum were included as fitted parameters. We found similar results to Sowell \etal\ \cite{Sowell++12aj}, but with errorbars typically slightly larger. Sowell \etal\ \cite{Sowell++12aj} used the Wilson-Devinney code \cite{WilsonDevinney71apj,Wilson79apj} and give no information on how their errorbars were obtained. Their errorbars are likely to be formal errors, which are known to be underestimated in many cases \cite{MaceroniRucinski97pasp,WilsonVanhamme04,PavlovskiMe09mn,Pavlovski+09mn,Pavlovski++18mn,Me+20mn}. Our best fits to the $y$ and $b$ data are shown in Fig.\,\ref{fig:sowell}.

%%%%%%%%%%%%%%%%%%%%%%%%%%%%%%%%%%%%%%%%%%%%%%%%%%%%%%%%%%%%%%%%%%%%%%%%%%%%%%%%%%%%%%%%%%%%%%%%%%%%%%%%%%%%%%%%%%%%%%%%%%%%%%%%%%%%%%%%%%%%%%%%%%%%%

\section*{Analysis of space-based light curve}

We then moved to modelling the TESS photometry, using only the data near eclipse. The orbital period and reference time of minimum were fitted, and we included the measured time of primary minimum from the analysis of the ground-based data above in order to increase the precision of the ephemeris. Quadratic functions were applied to the brightness of the system through each eclipse, to remove any remaining slow trends in brightness arising from either astrophysical or instrumental causes. We also retained the RVs of the two stars to help define the orbital shape of the system. Third light was checked for and found to be negligible, so was fixed at zero. The quality of the fit (Fig.\,\ref{fig:tess}) is good but not perfect, and the systematics in the residuals can be attributed to the presence of dark starspots on one or both stars (see below). The secondary eclipse occurs at an orbital phase of 0.6431.

The best fit of the TESS data occurs for a noticably larger ratio of the radii than for the $b$- and $y$-band ground-based data. To investigate this we ran a default solution, and used this to scale the errorbars of each individual dataset (TESS light curve and the RVs of each star) to force a reduced $\chi^2$ of unity. We then performed fits to these data with the ratio of the radii fixed at values from 0.9 to 1.1 in intervals of 0.002, and repeated this process for the $b$ and $y$ data. Fig.\,\ref{fig:kfix} shows the $\chi^2$ of the fits versus the ratio of the radii in all three cases. For the TESS data there is a clear minimum around $k=1$, whereas the $b$ and $y$ light curves have a much broader minimum around $k=0.96$. The TESS data clearly provide a more tightly constrained solution, and we attribute this to the complete coverage of all eclipse phases (the ground-based data miss the first and last contact points of secondary eclipse). Out of curiosity we repeated this process without the RVs, and found a negligible difference. We conclude that $e$ and $\omega$ are well defined by the photometry alone in this case.

With the results from the previous paragraph in mind, we proceeded to determine the physical parameters of KX\,Cnc using the TESS data and Sowell \etal\ \cite{Sowell++12aj} RVs, with errorbars scaled so each dataset yielded a reduced $\chi^2$ of unity. The systemic velocity was fitted separately for the two stars, but the values were found to be in good agreement. Uncertainties in the fitted parameters were calculated using Monte Carlo and residual-permutation algorithms \cite{Me08mn}. The final best fit and uncertainties are given in Table\,\ref{tab:lc}. Whilst the ratio of the radii is formally greater than unity, it is so by only an insignificant amount. The light ratio found in this solution agrees well with the spectroscopic value of 0.9685 found by Sowell \etal\ \cite{Sowell++12aj}.

\begin{table} \centering
\caption{\em Best fit to the TESS light curve and ground-based RVs of KX\,Cnc obtained with
{\sc jktebop}. The 1$\sigma$ uncertainties have been calculated using Monte Carlo and residual-permutation
algorithms. The same limb darkening coefficients were used for both stars. The uncertainties in the
systemic velocities do not account for any transformations onto a standard system, which are
likely much larger than the quoted errorbars. \label{tab:lc}}
\begin{tabular}{lr@{\,$\pm$\,}l}
{\em Parameter}                           & \multicolumn{2}{c}{\em Value}    \\[3pt]
{\it Fitted parameters:} \\
Primary eclipse time (BJD/TDB)            & 2458876.93684  & 0.00002         \\
Orbital period (d)                        &      31.2198786& 0.0000006       \\
Orbital inclination (\degr)               &      89.829    & 0.001           \\
Sum of the fractional radii               &       0.038580 & 0.000014        \\
Ratio of the radii                        &       1.0060   & 0.0039          \\
Central surface brightness ratio          &       0.9641   & 0.0015          \\
Linear limb darkening coefficient         &       0.2946   & 0.0047          \\
Quadratic limb darkening coefficient      & \multicolumn{2}{c}{0.21 (fixed)} \\
$e\cos\omega$                             &       0.20548  & 0.00004         \\
$e\sin\omega$                             &       0.42267  & 0.00042         \\
Velocity amplitude of star A (\kms)       &      50.021    & 0.095           \\
Velocity amplitude of star B (\kms)       &      50.485    & 0.053           \\
Systemic velocity of star A (\kms)        &       4.975    & 0.005           \\
Systemic velocity of star B (\kms)        &       5.032    & 0.004           \\[3pt]
{\it Derived parameters:} \\
Fractional radius of star A               &       0.01923  & 0.00010         \\
Fractional radius of star B               &       0.01935  & 0.00009         \\
Orbital eccentricity                      &       0.46997  & 0.00036         \\
Argument of periastron (\degr)            &      64.074    & 0.027           \\
Light ratio                               &       0.9756   & 0.0061          \\
\end{tabular}
\end{table}

%%%%%%%%%%%%%%%%%%%%%%%%%%%%%%%%%%%%%%%%%%%%%%%%%%%%%%%%%%%%%%%%%%%%%%%%%%%%%%%%%%%%%%%%%%%%%%%%%%%%%%%%%%%%%%%%%%%%%%%%%%%%%%%%%%%%%%%%%%%%%%%%%%%%%

\section*{Physical properties and distance}

For a full picture of the properties of KC\,Cnc we required estimates of the \Teff\ values of the stars. We obtained these from Sowell \etal\ \cite{Sowell++12aj}, and confirmed that their ratio was in good agreement with the central surface brightness ratio determined from the {\sc jktebop} solution of the TESS light curve. The \Teff\ values, and the surface gravities, of the stars are consistent with spectral types of G0\,V and G1\,V using the calibration by Pecaut \& Mamajek\cite{PecautMamajek13apjs}.

The \Teff\ values were augmented by the fractional radii, orbital inclination and eccentricity, period and velocity amplitude determined in the previous section, and provided to the {\sc jktabsdim} code \cite{Me++05aa,Me20obs}. This yielded the physical properties of the system, with uncertainties propagated using a perturbation approach, given in Table\,\ref{tab:absdim}. The masses and radii are determined to precisions of 0.5\% or better. The measured properties exhibit an inverted mass-radius relation -- star B is less massive but larger than star A -- which is not expected from stellar theory. However, this result is measured to only the 1.5$\sigma$ level (see the ratio of the radii in Table\,\ref{tab:lc}) so is not significant. The two stars are both physically very similar to our Sun.

\begin{table} \centering
\caption{\em Physical properties of KX\,Cnc. The \Teff\ values are from Sowell \etal\ \cite{Sowell++12aj}.
Units superscripted with an `N' are defined by IAU 2015 Resolution B3 \cite{Prsa+16aj}. \label{tab:absdim}}
\begin{tabular}{lr@{\,$\pm$\,}lr@{\,$\pm$\,}l}
{\em Parameter}        & \multicolumn{2}{c}{\em Star A} & \multicolumn{2}{c}{\em Star B} \\[3pt]
Mass ratio                                & \multicolumn{4}{c}{$0.9908 \pm 0.0021$}       \\
Semimajor axis (\Rsunnom)                 & \multicolumn{4}{c}{$54.744 \pm 0.060$}        \\
Mass (\Msunnom)                           &  1.1345 & 0.0032      &  1.1241 & 0.0045      \\
Radius (\Rsunnom)                         &  1.0527 & 0.0056      &  1.0593 & 0.0051      \\
Surface gravity ($\log$[cgs])             &  4.4825 & 0.0045      &  4.4388 & 0.0041      \\
Density ($\!$\rhosun)                     &   0.972 & 0.015       &   0.946 & 0.013       \\
Synchronous rotational velocity (\kms)    &   1.706 & 0.009       &   1.717 & 0.008       \\
\Teff\ (K)                                &    5900 & 100         &    5843 & 100         \\
Luminosity $\log(L/\Lsunnom)$             &   0.083 & 0.030       &   0.071 & 0.030       \\
$M_{\rm bol}$ (mag)                       &    4.53 & 0.07        &    4.57 & 0.08        \\
\end{tabular}
\end{table}

The distance to KX\,Cnc was determined using the apparent magnitudes of the system in the $BV$ and $JHK_s$ bands (see Table\,\ref{tab:info}) and the physical properties. This was done in two ways: the surface brightness method \cite{Me++05aa} with the empirical surface brightness calibration from Kervella \etal\ \cite{Kervella+04aa}, and the bolometric correction method with the theoretical bolometric corrections from Girardi \etal\ \cite{Girardi+02aa}. The distances found for the $BV$ bands are in good agreement with, but less precise than, the parallax distance of $49.30 \pm 0.12$\,pc from \textit{Gaia} DR2 \cite{Gaia18aa}. This agreement is evidence that the \Teff\ values of the stars are reliable, but it would be worthwhile in future to perform a detailed spectroscopic analysis in order to determine photospheric abundances and more precise \Teff\ values.

The distances obtained in the $JHK_s$ bands should be more reliable because of the tighter empirical calibration and lesser effect of interstellar extinction, but are anomalously large. On investigating this we found that the 2MASS observations of KX\,Cnc \cite{Cutri+03book} were taken at an orbital phase of 0.6410, which is during secondary eclipse (see Fig.\,\ref{fig:tess}) when the light from the system is fainter than the combined light of the two stars.

%%%%%%%%%%%%%%%%%%%%%%%%%%%%%%%%%%%%%%%%%%%%%%%%%%%%%%%%%%%%%%%%%%%%%%%%%%%%%%%%%%%%%%%%%%%%%%%%%%%%%%%%%%%%%%%%%%%%%%%%%%%%%%%%%%%%%%%%%%%%%%%%%%%%%

\section*{Starspot activity and tidal effects}

\begin{figure}[t] \centering \includegraphics[width=\textwidth]{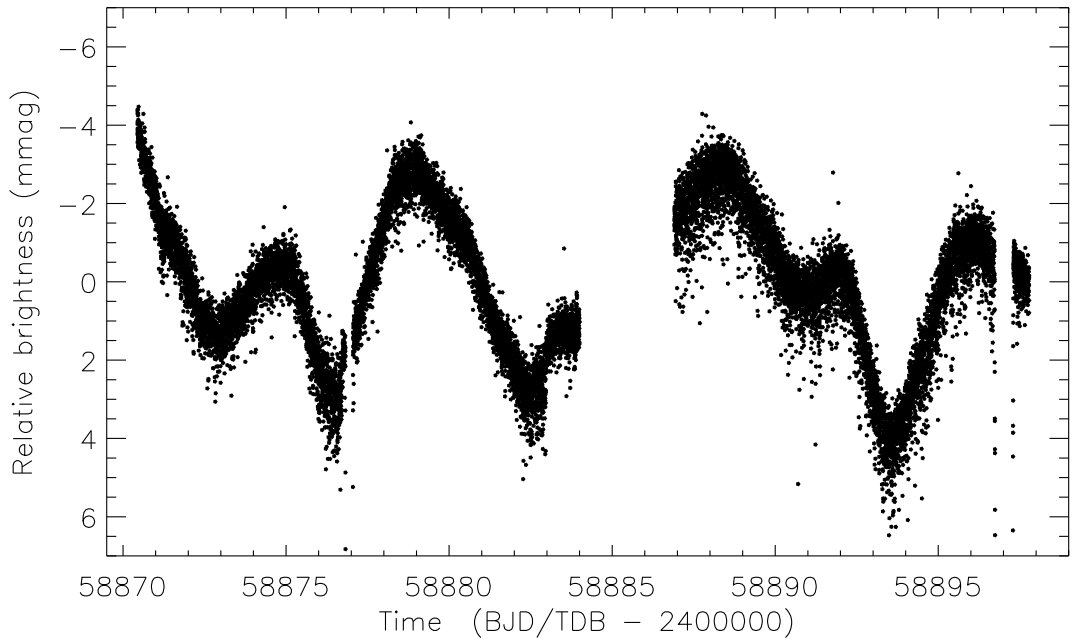} \\
\caption{\label{fig:spot} The TESS light curve of KX\,Cnc, with the y-axis limits chosen
to show the out-of-eclipse variability due to starspots.} \end{figure}

KX\,Cnc contains two solar-type stars so photometric variability due to starspots is a possibility \cite{Bouvier09conf,McQuillan++14apjs}. Fig.\,\ref{fig:spot} shows the TESS light curve of KX\,Cnc with a magnified y-axis to make the out-of-eclipse variability easier to see. There are clear signatures of rotational variability in the light curve, but evolution of the spots means that consecutive rotational periods of the star(s) do not repeat the same pattern of variability. A rotation period of $9.0 \pm 0.2$\,d provides a plausible solution to the rotational modulation seen in the TESS light curve of KX\,Cnc

Sowell \etal\ \cite{Sowell++12aj} measured a rotational period of 8.49\,d from photometric monitoring of the system over two observing seasons. This value is slightly shorter than suggested by the TESS light curve, and is based on data that are sparser but with a much better temporal baseline. Sowell \etal\ \cite{Sowell++12aj} measured rotational velocities $v\sin i$ of $6.4 \pm 1.0$\kms\ and $6.5 \pm 1.0$\kms\ for star A and star B, respectively, from their spectral line widths. With the radii measured in the previous section, these correspond to rotation periods of $8.3 \pm 0.1$\,d and $8.2 \pm 0.1$\,d, respectively.  The rotation of the stars is thus approximately consistent with the observed rotation periods, and is also greater than the synchronous and pseudosynchronous rotational velocities.

The system is therefore tidally unevolved: the orbit is not circularised and the stars are not rotating either synchronously or pseudosynchronously. This is not surprising due to the weakness of tidal effects at this orbital period. The theory of Zahn \cite{Zahn77aa} gives the timescales of orbital circularisation and rotational synchronisation to be approximately 10\,Gyr and 30\,Tyr, respectively (Zahn's equations 6.1 and 6.2 for convective-envelope stars). These are both significantly longer than the age of KX\,Cnc, in agreement with its eccentric orbit and supersynchronous stellar rotation \cite{Lurie+17aj}.

\section*{Conclusion}

The eclipsing binary system KX\,Cnc contains two stars similar to the Sun on an eccentric 31-d orbit. We have determined the physical properties of the system based on published RVs \cite{Sowell++12aj} and the TESS light curve, measuring masses and radii to precisions of 0.5\% or better. The TESS light curve is the first to have complete coverage of both eclipses, and leads to a ratio of the radii that is close to unity. It also shows clear brightness modulation due to starspots, and this modulation is consistent with the rotation period determined from ground-based light curves \cite{Sowell++12aj} and the spectroscopic rotational velocities of the stars \cite{Sowell++12aj}.

We have compared the masses, radii and \Teff\ values of the two stars to the predictions of several theoretical models \cite{Demarque+04apjs,Pietrinferni+04apj,Bressan+12mn}. This is relatively uninformative because the two stars are very similar, but does allow an age and chemical composition to be inferred. All properties of the stars can be matched for an age of 1.0--1.5\,Gyr and a metal abundance 1.5 times the solar value. This metal abundance matches the slightly super-solar metallicity inferred by Sowell \etal\ \cite{Sowell++12aj} from comparison between the spectra of KX\,Cnc and of standard stars.

Our understanding of KX\,Cnc would be improved by a detailed analysis of high-resolution spectra to obtain the \Teff\ values and photospheric chemical abundances of the stars. With this information, it will become a benchmark system capable of providing an important test of theoretical models of the evolution of solar-type stars.

%%%%%%%%%%%%%%%%%%%%%%%%%%%%%%%%%%%%%%%%%%%%%%%%%%%%%%%%%%%%%%%%%%%%%%%%%%%%%%%%%%%%%%%%%%%%%%%%%%%%%%%%%%%%%%%%%%%%%%%%%%%%%%%%%%%%%%%%%%%%%%%%%%%%%

\section*{Acknowledgements}

We acknowledge helpful discussions with Pierre Maxted. The following resources were used in the course of this work: the ESO archive; the NASA Astrophysics Data System; the SIMBAD database operated at CDS, Strasbourg, France; and the ar$\chi$iv scientific paper preprint service operated by Cornell University.

% \section*{Observatory style}
%
% Oxford ...ize
%
% Oxford and,
%
% Galileo {\em et al.}
%
% Smith \& Jones not `and'

% \bibliography{jkt}

%%%%%%%%%%%%%%%%%%%%%%%%%%%%%%%%%%%%%%%%%%%%%%%%%%%%%%%%%%%%%%%%%%%%%%%%%%%%%%%%%%%%%%%%%%%%%%%%%%%%%%%%%%%%%%%%%%%%%%%%%%%%%%%%%%%%%%%%%%%%%%%%%%%%%
\end{document}